# Efficient Personalized Web Mining: Utilizing The Most Utilized Data

L.K. Joshila Grace[1], V.Maheswari[2], Dhinaharan Nagamalai[3],
[1] Research Scholar, Department of Computer Science and Engineering
[2] Professor and Head, Department of Computer Applications
[1,2] Sathyabama University, Chennai, India
[3] Wireilla Net Solutions PTY Ltd, Australia

**Abstract.** Looking into the growth of information in the web it is a very tedious process of getting the exact information the user is looking for. Many search engines generate user profile related data listing. This paper involves one such process where the rating is given to the link that the user is clicking on. Rather than avoiding the uninterested links both interested links and the uninterested links are listed. But sorted according to the weightings given to each link by the number of visit made by the particular user and the amount of time spent on the particular link.

## 1 Introduction

While a search is being made for a particular word. Each user may have different views for a single word. For example if the search word is given as card. A child may be in need of some games in cards, an adult may want information regarding the ATM cards, an lay man would be in need of some ID card. So each person is in need of different information for a single word.

[4]Traditional approaches of mining user profiles are to use term vector spaces to represent user profiles and machine learning. [7]There are two kinds of method to learn users' interests. One method is the static method. The e-learning system sets questions or asks users to register, from which we can find out each user's information, such as age, major, courses to learn, courses learned, education background and so on. The other method is the dynamic method. The system observes each user's action through his session with the system. Then an analysis of his logs and queries are done to learn his interestingness.

Therefore the user profiling method was evolved. For performing search in a database or the web the user information like the name, address, occupation, Qualification, Area of interest etc are provided by the user. A separate data base table is maintained for the user. Whenever the user logs inside with his/her user name and password he would be able to do a personalized search. The search key given by the user would provide the list of links that are related. For the data file a separate database is present giving the keywords in the links. These key words will not include is, as, for, not etc. Only ten key words are considered. The search key matching the key words present in the database will get listed. The first time the user does the search he will get the normal list of links. The link that is clicked and used by the user is recorded in the database. So the next time the same user logs into the search engine and searches for the same search key the order of listing will change. The link that was clicked and used by the user would get the higher weightings, this gets into the first position in the list. Each time the search made by the user gets updated in the database. In this system not only the interested links but also the uninterested links gets displayed. But this may occupy the last few places in the list.

TextStat 3.0 software is used to find the key words in the document that has to be searched. This will give the frequently occurring words, omitting the is, was, the etc . this is the text where the match has to be found. These frequently occurring texts is called as the key word which are used to identify the particular link.

Incase of web usage mining the mining process is done based on the number of time the web site is visited and How much time spent on the web site[8][10][11].

- Preprocessing: Data preprocessing describes any type of processing performed on raw data to prepare it for another processing procedure. The different types of preprocessing in Web Usage Mining are:
    1. Usage Pre-Processing: Pre-Processing relating to Usage patterns of users.
    2. Content Pre-Processing: Pre- Processing of content accessed.
    3. Structure Pre-Processing: Pre-Processing related to structure of the website.
- Pattern Discovery: Data mining techniques have been applied to

Extract usage patterns from Web log data. The following are the pattern discovery
methods.
    1. Statistical Analysis
    2. Association Rules
    3. Clustering
    4. Classification
    5. Sequential Patterns
    6. Dependency

## 3 User Profiling

User profiling strategies can be broadly classified into two main approaches[2]:
    1. Document-based
    2. Concept-based approaches.

Document-based user profiling methods aim at capturing users' clicking and browsing behaviors. Users' document preferences are first extracted from the click through data, and then, used to learn the user behavior model which is usually represented as a set of weighted features.

On the other hand, concept-based user profiling methods aim at capturing users' conceptual needs. Users' browsed documents and search histories are automatically mapped into a set of topical categories. User profiles are created based on the users' preferences on the extracted topical categories.

## 4 Related works

There are many algorithms present which would list only the interested links of that particular user. Therefore if the user is in need of some more options he has to either log out and perform a normal search or has to go for any other process of searching.

### 4.1 Weighted association rule (WAR)

Each item in a transaction is assigned a weight to reflect the interest degree, which extends the traditional association rule method. [3]Weighted association rule (WAR) through associating a weight with each item in resulting association rules.

Each Web page is assigned to a weight according to interest degree and three key factors, i.e. visit frequency, stay duration and operation time.

### 4.2 PIGEON

Personalized Web page recommendation model called PIGEON abbr. for PersonalIzed web paGe rEcommendatiON) via collaborative filtering and a topic-aware Markov model. [5]A graph-based iteration algorithm to discover users' interested topics, based on which user similarities are measured.

### 4.3 Single Level Algorithm

This is a New pattern mining algorithm, for efficiently extracting approximate behavior patterns so that slight navigation variations can be ignored when extracting frequently occurring patterns.[6] This algorithm is particularly useful with websites that have a large number of web – pages.

### 4.4 FCA (formal concept analysis)

An approach that automatically mines web user profile based on FCA (formal concept analysis) methods from positive documents. [4]The formal concepts with their weights as patterns are represented to denote topics of interest. Based on the patterns discovered, the process of assessing documents relevance to the topics is found.

### 4.5 Other method

- **Joachims'** method assumes that a user would scan the search result list from top to bottom.[2] If a user has skipped a document di at rank i before clicking on document dj at rank j, it is assumed that he/she must have scan the document di and decided to skip it. Thus, we can conclude that the user prefers document dj more than document di
- **Click through method** - Click through data are important implicit feedback mechanism from users.[2] This gives the clicks made by the user on the same link or the different link.
- **Open Directory Project (ODP) -** [2]If a profile shows that a user is interested in certain categories, the search can be narrowed down by providing suggested results according to the user's preferred categories
- **Personalized query clustering method**
  This method involves these activities [2]
  1. Filtering the result by user profile
  2. Re-rank the retrievals from search engine

## 5 Tables generated

For maintaining all the information regarding the search various tables are created.

### 5.1 User profile

This table consists of the details that are provided by the user when they create a new account to use the search engine. The details may be user name, password, address, occupation, area of interest etc., These information's are stored in this table.

### 5.2 Users search keys

This table consists of the search key words used by the particular user, The option selected from the list. The number of times the search made by the user for the same search key, The number of times the selection of the same link made, the time spent on that particular link etc., are updated to the table. According to this information the weights for the links is put up. These weights would help in giving higher preference while listing the most interested links of the user in the next time of search.

### 5.3 Key words matching the document

This table consists of the key words of the document and the link of the document which has to be listed while searching for the particular search key is made by the user.

## 6 Search Process.

Every time the user logs inside the search engine. The login is verified with the username and password in the user profile database. Once the verification is successful it would link to the personalized search screen and the user can perform a search for a particular search key.

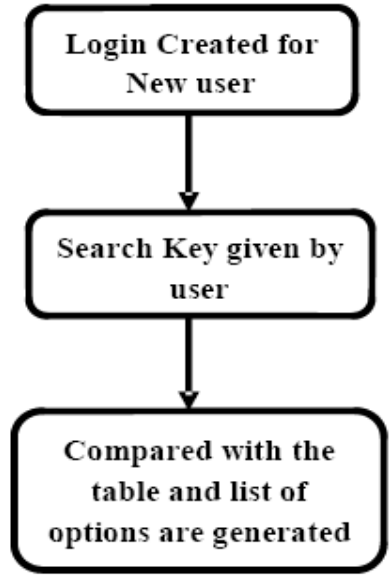

**Fig. 1.** Process for The first time of search

First step the search key is compared with the Key words matching the document table to analyze what are the links that has to be listed for the user. After the list of options (links) being listed the user activity is being listened by updating the details of the link he / she is clicking on and time duration they use the link data. This update is made in the user search keys table. The search key along with the link clicked by the user is noted in the table. The time duration the user is using the particular link is also found. This information would also contribute in providing the weights for each link. Not only the frequently visited link but also the highly utilized link is used. According to this information the weights for the links is put up. These weights would help in giving higher preference while listing the most interested links of the user in the next time of search.

The interested links include the frequently used and the amount of time spent on the particular link. This link occupies the first position in the list during the next time of search for the next time of search for the same search key. The fig.1 shows the procedure done while the user logs inside the search engine for the first time. Starts from the creation of login to the list of search link matching the search key

The next time when the same user logs in and performs a search for the same search key. The same process is carried out. The search key is compared with the key words of the document in the "Key words matching the document" table and listing option is generated. But since the user has already searched for the same key word the order of listing is differed according to the weightings generated from the previous search.

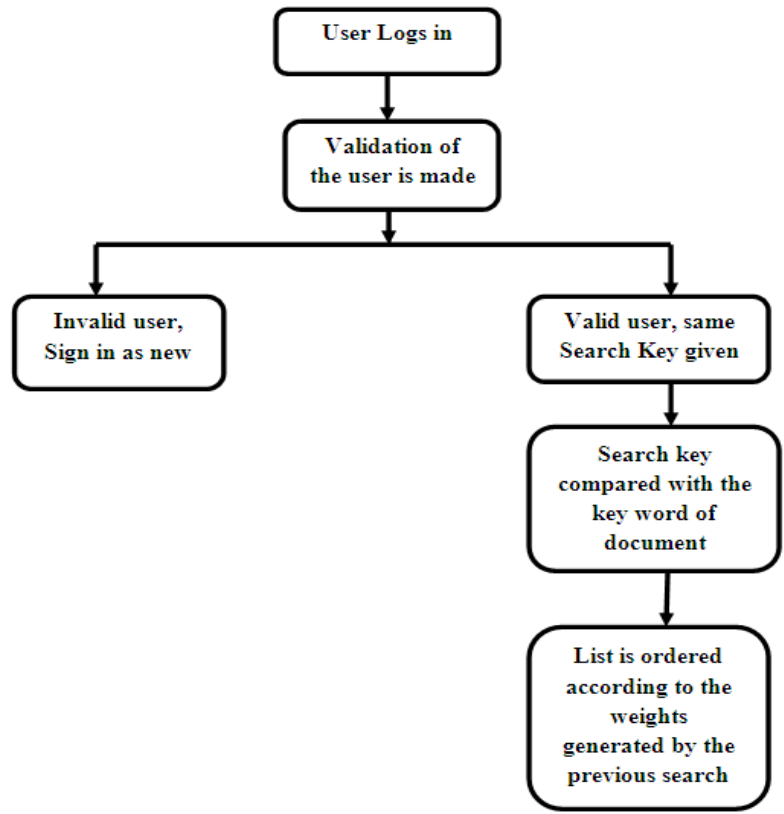

**Fig. 2.** Second time of search done by Same user for the same search key

The higher weighted link takes up the first position in the listing. The listing now will not only gives the interested link but also the uninterested link are listed.

## 7 Tools and Algorithm used

For developing these type of system various tools and algorithms are Used. They are discussed below

**7.1 Validation of User**

The user name and password is validated by comparing with the user Profile table. Only a valid user can perform this personalized search. Therefore the user who does not have a valid username and password has to sign in as a new user providing all the details. Then the user can continue with the personalized searching.

**7.2 Extracting key words from the documents**

TextStat 3.0 software tool is used to find the frequently used text omitting the words like is, was, are, the etc., [9] this tool provides information like

- Number of paragraphs:
- Number of words:
- Number of sentences:
- Number of printable characters (including spaces):
- Number of spaces:
- Number of tabulations:
- Number of Carriage Return:
- Number of Line Feed:
- Number of non-printable characters (others than the above):
- Number of words per sentence:

- Number of syllables per word (approximate):
- Flesch index:
- Start of list:

These are the information provided by the software. The last option gives the list of each word with the number of times repeated.

### 7.3 Identifying the weights

By default the user is specified with a weight of zero. Each time the user puts on a search key and clicks an option the weight of that particular link gets incremented. These weights provide the interest of the user. Thus the next time of search made by the user for the same search key would have the high weighted link in the first option of the list.

### 7.4 Extracting options from the weights

The lists of options are generated not only by just considering the link that match the key word but also the weights of the link. These weights are considered only for the second time of search by the user for the same key word.

The original GSP algorithm [1]
```
F1={frequent 1-sequences};
    For (k=2;Fk-1?Ø;k=k+1) do
        Ck=Set of candicate k-sequences;
        For all input-sequences ε in the databse do
            Increment count of all α Ck contained in ε
            Fk={α Ck | α.sup?min_sup};
        Set of frequent sequences = kFk
```

The evolved WTGSP algorithm [1]
```
WTGSP {
F1={frequent 1-sequences};
    For (k = 2; fk-1 ? Ø; k = k+1) do
        Ck = Set of candicate k-sequences;
        For all input-sequences ε in the databse do
            Increment count of all α Ck contained
            in ε With GETWeight(α)
            Fk = {α Ck | α.sup ? min_sup};
        Set of frequent sequences = kFk
        }
    Real function GetWeight(Date ItemDate) {
        int AllRecordsDistance = MaxDate - MinDate;
        int ItemDistance = ItemDate - MinDate;
        Real Weight = ItemDistance / AllRecordsDistance + 0.3;
        Return Weight;
        }
```

The Proposed WMGSP Algorithm
```
WMGSP {
F1={frequent 1-sequences};
    For (k = 2; fk-1 ? Ø; k = k+1) do
        Ck = Set of candicate k-sequences;
        For all input-sequences ε in the databse do
            Increment count of all α Ck contained
            in ε With GETWeight(α)

            Fk = {α Ck | α.sup ? min_sup};
        Set of frequent sequences = kFk
        }
```

```
Real function GetWeight(Date ItemDate) {
        int AllRecordsDistance = MaxDate - MinDate;
        int Minutes = startime – endtime;
        Real Weight = Minutes / AllRecordsDistance + 0.3;
        Return Weight;
}
```

By utilizing the weight generated by the Weight Minute GSP algorithm. The efficiency of determining the weights are even more refined by analyzing the utilization time period of the particular link.

## 8 Conclusion

In this system the utilization concept is provided for a definite set of data set. If the web server is connected and utilize a wide range of data it would be even more efficient. The time duration is calculated in terms of minutes if they are done in seconds it would give an even more fine output. A very effective mining of highly utilized data are listed to the user. This decreases the time of search and provides the exact information the user is in need of. The user can get a very effective personalized searching mechanism done. It uses the information of the utilization of the link by the user which helps to sort information for the same user during the next time of search.